\documentstyle[12pt]{article} 
\def \be{\begin{equation}} 
\def \ee{\end{equation}} 
\normalbaselines

\begin{document} 
\begin{flushright} 
TIFR/TH/97-22\\ 
15 May~1997
\end{flushright} 
\bibliographystyle{unsrt} 
\vskip0.5cm
\baselineskip=.8cm 
\begin{center} 
{ 
\LARGE \bf Factorization of the charge correlation function in 
$B^0\bar B^0$ oscillations
}
\\ 
[7mm] 
{ \bf G. V. Dass}$^1$ and {\bf K. V. L. Sarma}$^{2*}$ \\
$^1${\it Department of Physics, Indian Institute of Technology, 
Powai, Mumbai, 400 076, India }\\ 
$^2${\it Tata Institute of Fundamental Research, Homi Bhabha Road,
Mumbai, 400 005, India } 
\\ 
[10mm] 
\end{center}

\bigskip 

\begin{center} 
Abstract 
\end{center} 

\bigskip

\baselineskip=.8cm 

Extraction of the mass difference $\Delta m$ from $B^0\bar B^0$
oscillations involves tagging of bottom flavour at production and at
decay. We show that the asymmetry between the unmixed and mixed events
factorizes into two parts, one depending on the production-tag and the
other on the decay-tag.

\bigskip

\bigskip

\noindent PACS number(s): 14.40.Nd, 13.20.He, 11.30.Er.\\ 

\vfill

$^*$~E-mail: kvls@theory.tifr.res.in;~~fax: 091 22 215 2110 

\newpage 
\baselineskip=.8cm 

There is now a considerable body of experimental evidence for $B^0
\bar B^0$ oscillations. The general strategy adopted in gathering it
(for a review, see $e.g$., Ref. \cite{WU}) may be stated briefly as
follows: (A) identify the $B^0$ decay events that have one or more out
of a lepton $\ell ^{\pm }$, a $D^{*\mp }$, and a $K^{\pm }$, (B) tag
the bottom flavour at production by any convenient method: use of jet
charge or high-$p_T$ lepton in the opposite-hemisphere, or $B\pi
^{\pm}$ correlations in the same-side hemisphere, or asymmetry with
polarized electron beam, (C) measure the displacement of decay vertex
from the production vertex, (D) estimate the $B^0$ momentum, and (E)
convert the displacement into propagation time of the neutral
beon. From such measurements it has been demonstrated that the
mixed-events occur with a sinusoidal time-dependence that is
characteristic of oscillations. The frequency or mass-difference
$\Delta m$ of $B^0\bar B^0$ mixing has been extracted
\cite{AL93}-\cite{CDF} from the observed time-dependence.

Data on time-distribution are usually fitted to the charge correlation
function between the numbers of mixed and unmixed events 
\be 
C(t)= { N_{{\rm unmixed} }~ -~N_{{\rm mixed}} \over 
          N_{{\rm unmixed} }~ +~N_{{\rm mixed}} }~. 
                                                         \label{C} 
\ee 
Here $t$ is the decay time of $B^0$ measured
in its rest frame. The purpose of this note is to point out that
$C(t)$ factorizes into a part that depends on the production tag and
another that depends on the decay tag, provided we neglect terms that
contribute to second order of $CP$ violation. The time dependence of
$C(t)$ therefore will not be sensitive to the details and systematic
errors of the production tag.

Let us suppose that the time-distribution of $B^0$ decay is determined
by tagging a flavour-specific mode. Let the flavour at production be
determined by the jet charge of the $b$-jet in the opposite-side
hemisphere (jet charge is determined by a weighted sum of the charges
of the individual tracks; for details, see, $e.g.,$ Ref. \cite{OP94}).
We denote the probability to find a $b$-jet with normal jet charge
(=$-$1/3) by $P_n$. The jet from a $b$ can also have the abnormal
charge (=+1/3) if the $b$ fragmented into $\bar B^0$ or $\bar B_s^0$
which oscillated to the conjugate meson having positive bottom
flavour; let $P_a$ denote the probability to find a $b$-jet with
abnormal jet-charge. Hence the probabilities associated with jet
production are:
\be 
P_n={\rm Prob}\,(b \rightarrow J^-),~~P_a={\rm
Prob}\,(b \rightarrow J^{+}), 
\ee 
\be 
\bar P_n={\rm Prob}\,(\bar b \rightarrow J^{+}),~~\bar P_a={ \rm
Prob}\,(\bar b \rightarrow J^{-});
\ee 
here the superscript ($\pm $) on $J$ denotes the $sign$ of the jet
charge ($\pm 1/3$). Clearly, $CP$ invariance is violated if the
difference $(P_n-\bar P_n)$ or $(P_a-\bar P_a)$ is non-vanishing.

We take $t=0$ to be the instant at which the $b\bar b$ pair is
produced (we ignore the rare events with multiple $b\bar b$
pairs). Let the $\bar b$ fragmentation lead to a neutral beon $B^0$ or
$B_s^0$ which decays at time $t$ into a flavour-specific mode; this
could, for instance, be the fully-reconstructible mode $B^0
\rightarrow J/\psi~K^{*0}$ for studying $B^0\bar B^0$ oscillations, or
$B_s^0\rightarrow D_s^- ~\ell ^+~\nu $ for studying $B_s^0\bar B_s^0 $
oscillations \cite{ALbs}. In the following we focus on the inclusive
decay mode $B^0/\bar B^0 \rightarrow [D^*(2010)^{\pm } + {\rm
anything}]$ and define the decay rates into normal and abnormal modes
as
\be
D_n(t) = \Gamma (\bar B^0(t) \rightarrow
D^{*+}~+~ ...~),~~ D_a(t) = \Gamma (\bar B^0(t) \rightarrow D^{*-}~
+~...~).  
\ee 
\be 
\bar D_n(t) = \Gamma (B^0(t) \rightarrow D^{*-}~
+~...~),~~ \bar D_a(t) = \Gamma (B^0(t) \rightarrow D^{*+}~+~...~).  
\ee
Here, the dots $(~...~)$ indicate `anything'; $ \bar B^0(t) $ is the
physical state which evolved from a $\bar B^0$ state after a lapse of 
time $t$; the decay rates
corresponding to initial antiquark ($\bar b$) have a
`bar' on them. Decays classified as abnormal are due to $B^0\bar B^0$
oscillations. The conditions implied by $CP$ invariance are
$[D_n(t)-\bar D_n(t)]=0$ and $[D_a(t)-\bar D_a(t)]=0$.

We now write the relative numbers of events $N({ji})\,$, where the
first sign $j$ is the sign of the jet charge and the second
sign $i$ is the sign of the $D^*$ charge:
\begin{eqnarray} 
N({--}) &=& P_n\,\bar D_n + \bar P_a\, D_a~, \label{enf}\\ 
N({++}) &=& \bar P_n\, D_n + P_a\, \bar D_a~,\\ 
N({+-}) &=& \bar P_n\, D_a + P_a \,\bar D_n~, \\ 
N({-+}) &=& P_n\, \bar D_a + \bar P_a \,D_n~. \label{enl}  
\end{eqnarray} 
These expressions are easily interpreted: for instance, $N({+-})$
refers to events with positive jet charge arising from the decay of
either a $\bar b$-hadron which did not oscillate (called normal), or a
$b$-hadron which did oscillate (called abnormal); the $D^{*-}$ in the
measurement-hemisphere can result from the decay following either the
transition $B^0(t)\rightarrow B^0$ or the transition $\bar
B^0(t)\rightarrow B^0$.

The relative number of unmixed events containing a bottom jet and the
inclusive decay $B^0/\bar B^0\rightarrow D^{*\pm }~+ {\rm anything} $,
is given by
\begin{eqnarray}
 N_{{\rm unmixed}}
&=& N({--})\,+ N({++}) \label{unmix} \\
                  &=& {1 \over 2}[\,(P_n+\bar P_n)(D_n+\bar D_n )+
                   (P_a+\bar P_a)(D_a+\bar D_a ) \nonumber \\ 
                  &~&~~~- (P_n-\bar P_n)(D_n-\bar D_n )- 
                     (P_a-\bar P_a)(D_a-\bar D_a )\,] \\ 
                 &\simeq & {1 \over 2}[\,(P_n+\bar P_n)(D_n+
                   \bar D_n)+ (P_a+\bar P_a)(D_a+\bar D_a )\,].
                 \label{app1} 
\end{eqnarray} 
The last step neglects terms that contribute to second order of $CP$
violation. The number of mixed events (namely, events having abnormal
charge either at production or at decay, but not both) is similarly 
given by
\begin{eqnarray} 
N_{{\rm mixed}} &=& N({+-})\, + \, N({-+}) \label{mix}\\
                  &=& {1 \over 2}[\,(P_n+\bar P_n)(D_a+\bar D_a )+
                   (P_a+\bar P_a)(D_n+\bar D_n ) \nonumber \\ 
                  &~&~~~- (P_n-\bar P_n)(D_a-\bar D_a )- 
                     (P_a-\bar P_a)(D_n-\bar D_n )\,] \\ 
               &\simeq & {1\over 2}[\,(P_n+\bar P_n)(D_a+\bar D_a)+ 
                       (P_a +\bar P_a)(D_n+\bar D_n)\,],
                       \label{app2} 
\end{eqnarray} 
wherein the neglected terms of second-order of $CP$-violation are
different from those neglected in Eq. (\ref{app1}). Note that in
Eqs. (\ref{unmix}) and (\ref{mix}) the charge labels in
$N(ji)$ depend on the tags used for production and decay.

The charge-correlation function $C(t)$ is $CP$-even. It contains the
products $ (P_i-\bar P_i)(D_j-\bar D_j)$, with $(i,j)=(n,a)$, which,
being quadratic in $CP$-violation, are presumably small (the
mass-matrix $CP$ violation is believed anyway to be too small to be 
relevant). Hence it is quite reasonable to substitute Eqs.
(\ref{app1}) and (\ref{app2}) in Eq. (\ref{C}). Thus $C$ takes the
factorized form
\be 
C(t) \simeq ~{\left[ (P_n+\bar P_n) - (P_a+\bar P_a)
               \over 
                  (P_n +\bar P_n) + (P_a +\bar P_a)\right]}
              \times {\left[ (D_n+\bar D_n) - (D_a+\bar D_a) 
               \over 
                   (D_n +\bar D_n) + (D_a +\bar D_a)\right]}~.  
                                                   \label{main} 
\ee
The first (time-independent) factor involving $P$'s refers to the
production tag and the second involving $D(t)$'s to the decay tag.
Obviously Eq. (\ref{main}) will be exact if the $CP$-violating
differences $(P_i-\bar P_i)$ or $(D_i-\bar D_i)$ vanish for $i=n$ 
and $i=a$.

Factorization of $C(t)$ is valid under general conditions: There is no
need to consider the production tag in a time-integrated version which
is inherent to the jet-charge method; the double-time distribution
also will factorize. As for decay, any specific channel that can tag
the bottom will suffice.

When a lepton tag is used either at the production end {\it or} at the
decay end, factorization would automatically follow. This is because
of the general result (see, $e.g.,$ Ref. \cite{BELL}) that direct $CP$
violation cannot show up in decay channels which include a lepton pair
$\ell \nu $, provided we assume $CPT$-invariance, retain terms of
lowest-order of the weak Hamiltonian and ignore the electroweak
scattering phase-shifts. Recently the ALEPH group \cite{AL96} has
verified by simulations that the `lepton-signed jet charge' $Q_{\ell
H}(t)$ does factorize. This is to be expected because, if we now let
the label $i$ in $N(ji)$ denote the lepton charge, the negative
average of the product charges is given by
\begin{eqnarray} Q_{\ell H}(t) 
&=&
-~ {\sum_{j}\sum_{i}~(j)(i)~N(ji) \over \sum_{j}\sum_{i}~N(ji) }\\ 
&=& C(t).  
\end{eqnarray}

When the production-tag and the decay-tag are non-leptonic (for
instance, $K^{\pm }$ and opposite-side jet, as in Ref. \cite{DE96}),
the terms that prevent factorization can be neglected as they involve
second-order of $CP$ violation. There is yet another case in which
Eq. (\ref{main}) is exact: this is when the production tag originates
from a $CP$-invariant interaction. Examples of experimental interest
are the asymmetry with polarized electrons \cite{SLD} based on
neutral-current electroweak interaction and the $B\pi ^{\pm}$
correlations \cite{CDF} based on strong interaction.

On the other hand, it may be noted that the function
\be 
\chi (t) = 
{N_{{\rm mixed}} \over N_{{ \rm unmixed} }~ +~
                                       N_{{\rm mixed}}}~
         = {1-C(t)\over 2}~ 
\ee 
(which is akin to the dilepton mixing ratio), does not factorize; the
ratio ($N_{\rm mixed}/ N_{\rm unmixed}$) also does not factorize.

In conclusion, when a pair of bottom particles is incoherently
produced (as in $Z$ decays), the asymmetry $C(t)$ between the unmixed
and mixed events is well-suited for studying $B^0\bar B^0$ or $B_s^0
\bar B_s^0$ oscillations. It separates into two factors, one depending
on the production-tag and the other on the decay-tag. For this reason,
sensitivity to systematic errors associated with the production tag
would be minimal if the frequency $\Delta m$ is extracted by using
$C(t)$.


\newpage
\begin{thebibliography}{\mbox{ }} 

\bibitem{WU} S.L. Wu, Talk at the 17th International Symposium on
Lepton-Photon Interactions at High Energies, Beijing, 1995, CERN
preprint CERN-PPE/96-82 (1996).

\bibitem{AL93} ALEPH Collaboration, D. Buskulic {\it et~al.,}
Phys. Lett. B {\bf 313}, 498 (1993); {\bf 322}, 441 (1994).

\bibitem{AL96} ALEPH Collaboration, D. Buskulic {\it et~al.,} CERN
preprint CERN-PPE/96-102, July 1996, Submitted to Z. Phys. C.

\bibitem{OP94} OPAL Collaboration, R. Akers {\it et~al.,} Phys. Lett.
B {\bf 327}, 411 (1994).

\bibitem{OP94a} OPAL Collaboration, R. Akers {\it et~al.,}
Phys. Lett. B {\bf 336}, 585 (1994); Z. Phys. C {\bf 66}, 555 (1995);
G. Alexander {\it et~al.,} Z. Phys. C {\bf 72}, 377 (1996).

\bibitem{DE94} DELPHI Collaboration, P. Abreu {\it et~al.,} Phys.
Lett. B {\bf 338}, 409 (1994).

\bibitem{DE96} DELPHI Collaboration, P. Abreu {\it et~al.,} Z. Phys. C
{\bf 72}, 17 (1996).

\bibitem{L3} L3 Collaboration, M. Acciarri {\it et~al.,} Phys.
Lett. B {\bf 383}, 487 (1996).

\bibitem{SLD} SLD Collaboration, K. Abe {\it et~al.,} Reports at the
XXVIII International Conference on High Energy Physics, July 1996,
Warsaw, Poland: SLAC-PUB-7228,7229,7230 (July 1996).

\bibitem{CDF} CDF Collaboration, Reported at the XXVIII International
Conference on High Energy Physics, July 1996, Warsaw, Poland: Fermilab
preprint FERMILAB-Conf-96/175-E (Sept 1996).

\bibitem{ALbs} ALEPH Collaboration, D. Buskulic {\it et~al.,}
Phys. Lett. B {\bf 377}, 205 (1996). 

\bibitem{BELL} J.S. Bell, in {\it High Energy Physics}, Les Houches
1965, edited by C. DeWitt and M. Jacob, (Gordon and Breach, 1965),
p. 403; P.K. Kabir, in {\it Particle Interactions at High Energies},
edited by T.W. Priest and L.L.J. Vick, (Oliver \& Boyd, 1967), p. 248.

\end{thebibliography}
\end{document}